\documentclass[english]{article}
\usepackage{lmodern}

\usepackage[T1]{fontenc}
\usepackage[latin9]{inputenc}
\usepackage{amsmath}
\usepackage{graphicx}
\usepackage{esint}
\usepackage[authoryear]{natbib}

\makeatletter

\providecommand{\tabularnewline}{\\}

\usepackage{natbib,alifeconf}
\bibpunct{(}{)}{,}{a}{,}{,}

\author {Artemy Kolchinsky$^{1,2}$ and Luis M. Rocha$^{1,2}$ \\
\mbox{} \\
$^1$ School of Informatics and Computing, Indiana University, Bloomington IN 47401, USA \\
$^2$ FLAD Computational Biology Collaboratorium, Instituto Gulbenkian de Ci\^{e}ncia, Portugal \\
\{akolchin,rocha\}@indiana.edu}

\let\originalleft\left
\let\originalright\right
\def\left#1{\mathopen{}\originalleft#1}
\def\right#1{\originalright#1\mathclose{}}

\makeatother

\usepackage{babel}
\begin{document}

\title{Prediction and Modularity in Dynamical Systems}
\maketitle
\begin{abstract}
Identifying and understanding modular organizations is centrally important
in the study of complex systems. Several approaches to this problem
have been advanced, many framed in information-theoretic terms. Our
treatment starts from the complementary point of view of statistical
modeling and prediction of dynamical systems. It is known that for
finite amounts of training data, simpler models can have greater predictive
power than more complex ones. We use the trade-off between model simplicity
and predictive accuracy to generate optimal multiscale decompositions
of dynamical networks into weakly-coupled, simple modules. State-dependent
and causal versions of our method are also proposed.
\end{abstract}

\section{Introduction}

The study of complex dynamical systems -- such as gene regulatory
networks \citep{han2004evidence}, structural and functional brain
networks \citep{bullmore2009complex}, ecological food webs \citep{krause2003compartments},
and others \citep{hartwell1999molecular,schlosser2004modularity}
-- has frequently uncovered the presence of modularity. Broadly speaking,
modular systems are composed of tightly-integrated subsystems, called
\emph{modules,} which are in turn weakly coupled to one another.

Numerous explanations have been proposed for the function of modularity
in complex systems, only a few of which are mentioned here. \citet{simon1962architecture}
suggested that modularity can contain the effects of harmful perturbations
and lead to greater developmental and operational robustness, especially
when modules are hierarchically arranged. \citet{kashtan2005spontaneous}
argued that modular systems can take advantage of reusability when
adapting to changing combinations of fixed environmental tasks. \citet{tononi1998complexity}
proposed that modularity balances the conflicting needs for subsystems
that are functionally specialized but also integrated into globally
coherent states. Notably, it has also been shown to arise as a result
of non-adaptive processes, such as neutral evolution of gene regulatory
networks \citep{force2005origin,sole2008spontaneous} and stochastic
fluctuations in network connectivity patterns \citep{guimera2004modularity}.

Though the concept of modularity has acquired a central place in the
study of complex systems, its meaning and operationalization varies
widely between scientific paradigms, fields, and processes of interest.
In the biological sciences alone, one can find references to \emph{structural},
\emph{developmental}, \emph{physiological}, \emph{variational}, and
\emph{functional} \emph{modularity} \citep{winther2001varieties,wagner2007road},
among others. In this work, we propose a formal notion of modularity
based on statistical modeling. Our approach applies to a broad class
of discrete-time multivariate dynamics, whether represented by dynamic
models, such as Boolean or dynamic Bayesian networks, or empirical
distributions estimated from time series recordings. Unlike much recent
work on community-structure in static graphs, we identify modularity
in the organization of dynamically interacting components. We argue
that in addition to being useful for analysis of real-life dynamical
systems, our approach can shed light on connections between notions
of modularity utilized in different domains, as well as the general
role of modularity in modeling.

The next section provides a brief background on information theory.
We then outline traditional information-theoretic approaches to modularity
in dynamical systems, and develop our own treatment in terms of statistical
modeling. After applying it to an example dynamical system, we consider
state-dependent and causal versions of modular decompositions. We
conclude by discussing issues of parameterization, directions for
further work, and connections between our method and broader questions
of modeling.

\section{Information theory}

Information theory provides principled measures of information transfer
and statistical dependence in distributed systems. As such, it is
well-suited for quantifying measures of coupling and modularity.

To review, Shannon \emph{entropy }measures the uncertainty in the
measurement outcomes of a random variable. If $X$ is a discrete random
variable with an associated probability distribution $P\left(X\right)$,
then its entropy is:
\[
\ensuremath{H\left(X\right)=-\sum_{x\in X}P\left(x\right)\log P\left(x\right)}
\]
A random variable that takes a single value with probability 1 has
an entropy of 0, while an equiprobable random variable assumes the
maximum entropy of $\log\;\left|X\right|$, where $\left|X\right|$
is the number of possible outcomes. When the base of the logarithm
is 2, as in this work, the units of entropy are \emph{bits} (1 bit
is the uncertainty in the choice between 2 equally possible outcomes).
Because measuring a variable reduces uncertainty about its value,
entropy can also be considered a measure of information. 

When provided with a joint distribution over two random variables
such as $P\left(X,Y\right)$, \emph{conditional entropy} measures
the expected uncertainty in the value of one variable given that the
value of the other is known:
\[
H\left(X\vert Y\right)=H\left(X,Y\right)-H\left(Y\right)=-\sum_{x,y}P\left(x,y\right)\log P\left(x\middle\vert y\right)
\]

\emph{Mutual information} is a symmetric measure of nonlinear correlation
between two random variables. Expressed as the difference between
entropy and conditional entropy, it can be interpreted as the reduction
in uncertainty about the value of one random variable provided by
knowledge of the other:
\begin{eqnarray*}
I\left(X;Y\right) & = & H\left(X\right)+H\left(Y\right)-H\left(Y,X\right)\\
 & = & H\left(X\right)-H\left(X\middle\vert Y\right)=H\left(Y\right)-H\left(Y\middle\vert X\right)\\
 & = & \sum_{x,y}P\left(x,y\right)\log\frac{P\left(x,y\right)}{P\left(x\right)P\left(y\right)}
\end{eqnarray*}
Mutual information captures the amount of constraint in the joint
distribution of two variables not present in their marginal distributions.
It is equal to 0 when two variables are statistically independent,
and reaches its maximum possible value of $\min\left\{ H\left(X\right),H\left(Y\right)\right\} $
when one variable is a function of the other.

Mutual information can be extended to the case of more than two variables.
Let random vector $\mathbf{X}$=$\left(X_{1},X_{2},\dots,X_{L}\right)$
with distribution $P\left(\mathbf{X}\right)$ represent the state
of a system composed of $L$ distinct variables. The total constraint
in this system not present in any single variable is measured by a
multivariate version of mutual information, often called \emph{multi-information}
\citep{studeny1998multiinformation} or \emph{integration} \citep{tononi1994measure}:
\begin{eqnarray}
\mathcal{I}\left(\mathbf{X}\right) & = & \sum_{i=1}^{L}H\left(X_{i}\right)-H\left(\mathbf{X}\right)\label{eq:multiinformation}\\
 & = & \sum_{\mathbf{x}}P\left(\mathbf{x}\right)\log\frac{P\left(\mathbf{x}\right)}{\prod_{i=1}^{L}P\left(x_{i}\right)}\nonumber 
\end{eqnarray}

\emph{Kullback-Leibler (KL) divergence }is a measure of the difference
between two distributions:
\begin{equation}
\mathrm{KL}\left(P\middle\Vert Q\right)=\sum_{x}P\left(x\right)\log\frac{P\left(x\right)}{Q\left(x\right)}\label{eq:kl}
\end{equation}
It is always positive and $0$ iff $P=Q$, though it is not a distance
because it is not symmetric. Importantly, many information-theoretic
measures can be restated in terms of KL divergence. For example, the
multi-information of eq. \ref{eq:multiinformation} is equal to the
KL divergence between the distribution of $\mathbf{X}$ and a product
of the marginal distributions over the individual variables of $\mathbf{X}$.

\section{Modularity in multivariate dynamics}

As previously mentioned, multi-information measures the total amount
of higher-order constraint present among the variables of a multivariate
system. It is 0 when these variables are independent, and increases
when more statistical interaction between variables is present \citep{studeny1998multiinformation}.
For this reason, many formal approaches to modularity search for system
transformations that minimize this measure.

Several kinds of transformations can be investigated. \emph{Independent
component analysis} attempts to minimize multi-information over the
space of linear mappings (coordinate changes) of a multivariate system
\citep{hyvarinen2000independent}. A different approach, closer to
the one pursued here, looks for\emph{ partitions} of system variables
with low multi-information.

A partition $\pi$ of set $S$ is a set of mutually exclusive, nonempty
subsets $B\subseteq S$, called \emph{blocks}, such that $\bigcup_{B\in\pi}B=S$.
For example, $\left\{ \left\{ 1\right\} ,\left\{ 2,3\right\} \right\} $
and $\left\{ \left\{ 1,2,3\right\} \right\} $ are two possible partitions
of the set $\left\{ 1,2,3\right\} $. We also use a more concise notation:
the two partitions above, for example, can be referred to as 1/23
and 123 respectively. Additionally, $\pi_{0}$ is used to indicate
the \emph{total partition}, which includes the entire set in a single
block, i.e. $\pi_{0}\equiv\left\{ S\right\} $.

We look at partitions of $V=\left\{ 1,\dots,L\right\} $, the set
of indexes of the variables of random vector $\mathbf{X}$. For partition
$\pi$ and block $B\in\pi$, $P\left(\mathbf{X}_{B}\right)$ indicates
the marginalization of $P\left(\mathbf{X}\right)$ onto the variables
whose indexes are in $B$. For example, $P\big(\mathbf{X}_{\left\{ 1,2\right\} }\big)$
is the marginal distribution of the first two variables of $\mathbf{X}$.

We define the multi-information of partition $\pi$ as: 
\[
\mathcal{I}_{\pi}\left(\mathbf{X}\right)=\sum_{B\in\pi}H\left(\mathbf{X}_{B}\right)-H\left(\mathbf{X}\right)
\]
This measure quantifies the amount of constraint holding among the
blocks of $\pi$. Finding partitions with low multi-information corresponds
to identifying weakly-coupled subsystems. Variations on this theme
appear in information-theoretic treatments of modularity starting
from early cybernetics \citep{conant1972dsc} to more recent approaches
in computational neuroscience \citep{tononi2003measuring}.

Multi-information is defined over a time-invariant distribution of
system states. Though it does not account for the dynamic flow of
information within a system, it can be generalized to this case. Assume
a multivariate system with Markovian dynamics represented by $P\left(\mathbf{X}^{\prime}=\mathbf{x}^{\prime}\middle\vert\mathbf{X}=\mathbf{x}\right)$,
the conditional probability distribution of transitioning to each
future state $\mathbf{x}^{\prime}$ given starting state $\mathbf{x}$,
as well as $P\left(\mathbf{X}=\mathbf{x}\right)$, the distribution
over starting states.%
\footnote{We assume that the dynamics are stationary, in that the transition
probability distribution does not change through time. Our analysis
can also be applied to higher-order Markovian systems, though for
simplicity they are not considered here.%
} The amount of information flowing dynamically among the blocks of
$\pi$ is called \emph{stochastic interaction} \citep{ay2003dynamical}.
It is a conditional version of KL divergence between the transition
distribution of the whole system and the product of marginal transition
distributions of the variable blocks specified by partition $\pi$:
\begin{eqnarray}
\mathcal{I}_{\pi}\left(\mathbf{X}^{\prime}\middle\vert\mathbf{X}\right) & = & \sum_{B\in\pi}H\left(\mathbf{X}_{B}^{\prime}\middle\vert\mathbf{X}_{B}\right)-H\left(\mathbf{X}^{\prime}\middle\vert\mathbf{X}\right)\label{eq:stochastic_interaction}\\
 & = & \mathrm{KL}\left[P\left(\mathbf{X}^{\prime}\vert\mathbf{X}\right)\middle\Vert{\displaystyle \prod_{B\in\pi}}P\left(\mathbf{X}_{B}^{\prime}\middle\vert\mathbf{X}_{B}\right)\right]\nonumber 
\end{eqnarray}
These kinds of dynamic generalizations of multi-information have recently
been proposed as measures of system-wide coupling in brain dynamics
\citep{balduzzi2008integrated,barrett2011practical}. 

\begin{figure}
\begin{minipage}[t]{0.35\columnwidth}%
\hfill{}

\includegraphics[clip,scale=0.64]{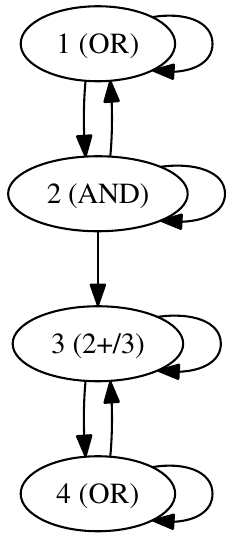}%
\end{minipage}%
\begin{minipage}[t]{0.3\columnwidth}%
{\tiny \hfill{}}{\tiny \par}

{\tiny }%
\begin{tabular}{cc}
\textsf{\tiny $\substack{\textsf{\textsf{Starting}}\\
\textsf{state}
}
$} & \textsf{\tiny $\substack{\textsf{Future}\\
\textsf{state}
}
$}\tabularnewline
\hline 
\noalign{\vskip-0.04cm}
\textsf{\tiny 0000} & \textsf{\tiny 0000}\tabularnewline[-0.04cm]
\noalign{\vskip-0.04cm}
\textsf{\tiny 0001} & \textsf{\tiny 0001}\tabularnewline[-0.04cm]
\noalign{\vskip-0.04cm}
\textsf{\tiny 0010} & \textsf{\tiny 0001}\tabularnewline[-0.04cm]
\noalign{\vskip-0.04cm}
\textsf{\tiny 0011} & \textsf{\tiny 0011}\tabularnewline[-0.04cm]
\noalign{\vskip-0.04cm}
\textsf{\tiny 0100} & \textsf{\tiny 1000}\tabularnewline[-0.04cm]
\noalign{\vskip-0.04cm}
\textsf{\tiny 0101} & \textsf{\tiny 1011}\tabularnewline[-0.04cm]
\noalign{\vskip-0.04cm}
\textsf{\tiny 0110} & \textsf{\tiny 1011}\tabularnewline[-0.04cm]
\noalign{\vskip-0.04cm}
\textsf{\tiny 0111} & \textsf{\tiny 1011}\tabularnewline[-0.04cm]
\noalign{\vskip-0.04cm}
\textsf{\tiny 1000} & \textsf{\tiny 1000}\tabularnewline[-0.04cm]
\noalign{\vskip-0.04cm}
\textsf{\tiny 1001} & \textsf{\tiny 1001}\tabularnewline[-0.04cm]
\noalign{\vskip-0.04cm}
\textsf{\tiny 1010} & \textsf{\tiny 1001}\tabularnewline[-0.04cm]
\noalign{\vskip-0.04cm}
\textsf{\tiny 1011} & \textsf{\tiny 1011}\tabularnewline[-0.04cm]
\noalign{\vskip-0.04cm}
\textsf{\tiny 1100} & \textsf{\tiny 1100}\tabularnewline[-0.04cm]
\noalign{\vskip-0.04cm}
\textsf{\tiny 1101} & \textsf{\tiny 1111}\tabularnewline[-0.04cm]
\noalign{\vskip-0.04cm}
\textsf{\tiny 1110} & \textsf{\tiny 1111}\tabularnewline[-0.04cm]
\noalign{\vskip-0.04cm}
\textsf{\tiny 1111} & \textsf{\tiny 1111}\tabularnewline[-0.04cm]
\end{tabular}%
\end{minipage}%
\begin{minipage}[t]{0.3\columnwidth}%
\hfill{}

\begin{tabular}{cc}
\multicolumn{2}{r}{\textsf{\footnotesize $\;\;\;\pi\;\;\;\;\mathcal{I}_{\pi}\left(\mathbf{X}'\middle\vert\mathbf{X}\right)$}}\tabularnewline
\hline 
\noalign{\vskip-0.03cm}
\textsf{\footnotesize 1234} & \textsf{\footnotesize 0.00}\tabularnewline[-0.03cm]
\noalign{\vskip-0.03cm}
\textsf{\footnotesize 12 / 34} & \textsf{\footnotesize 0.50}\tabularnewline[-0.03cm]
\noalign{\vskip-0.03cm}
\textsf{\footnotesize 1 / 234} & \textsf{\footnotesize 1.00}\tabularnewline[-0.03cm]
\noalign{\vskip-0.03cm}
\textsf{\footnotesize 123 / 4} & \textsf{\footnotesize 1.00}\tabularnewline[-0.03cm]
\noalign{\vskip-0.03cm}
\textsf{\footnotesize 134 / 2} & \textsf{\footnotesize 1.25}\tabularnewline[-0.03cm]
\noalign{\vskip-0.03cm}
\textsf{\footnotesize 124 / 3} & \textsf{\footnotesize 1.31}\tabularnewline[-0.03cm]
\noalign{\vskip-0.03cm}
\textsf{\footnotesize 12 / 3 / 4} & \textsf{\footnotesize 1.31}\tabularnewline[-0.03cm]
\noalign{\vskip-0.03cm}
\textsf{\footnotesize 1 / 2 / 34} & \textsf{\footnotesize 1.50}\tabularnewline[-0.03cm]
\noalign{\vskip-0.03cm}
\textsf{\footnotesize 14 / 23} & \textsf{\footnotesize 2.00}\tabularnewline[-0.03cm]
\noalign{\vskip-0.03cm}
\textsf{\footnotesize 1 / 23 / 4} & \textsf{\footnotesize 2.00}\tabularnewline[-0.03cm]
\noalign{\vskip-0.03cm}
\textsf{\footnotesize 13 / 24} & \textsf{\footnotesize 2.16}\tabularnewline[-0.03cm]
\noalign{\vskip-0.03cm}
\textsf{\footnotesize 13 / 2 / 4} & \textsf{\footnotesize 2.16}\tabularnewline[-0.03cm]
\noalign{\vskip-0.03cm}
\textsf{\footnotesize 14 / 2 / 3} & \textsf{\footnotesize 2.31}\tabularnewline[-0.03cm]
\noalign{\vskip-0.03cm}
\textsf{\footnotesize 1 / 24 / 3} & \textsf{\footnotesize 2.31}\tabularnewline[-0.03cm]
\noalign{\vskip-0.03cm}
\textsf{\footnotesize 1 / 2 / 3 / 4} & \textsf{\footnotesize 2.31}\tabularnewline[-0.03cm]
\end{tabular}%
\end{minipage}\caption{A simple four node Boolean network (nodes 1, 2, 3, and 4 perform OR,
AND, majority, and OR update functions respectively). Its full state
transition table is shown in center. On the right, the stochastic
interaction of every possible partition of the network.\label{fig:Simple-four-node}}
\end{figure}

A simple demonstration is provided by the Boolean network in fig.
\ref{fig:Simple-four-node}. It has four nodes, whose update functions
are OR, AND, majority rule, and OR respectively. The stochastic interaction
of each possible partition is provided, assuming a uniform distribution
over starting states. For example, the partition 12/34 is the bi-partition
having the lowest stochastic interaction: the block $\left\{ 1,2\right\} $
has conditional entropy $H\big(\mathbf{X}_{\left\{ 1,2\right\} }^{\prime}\vert\mathbf{X}_{\left\{ 1,2\right\} }\big)=0$
(nodes 1 and 2 do not depend on the rest of the system, so their marginalized
dynamics are deterministic), while block $\left\{ 3,4\right\} $ has
conditional entropy $H\big(\mathbf{X}_{\left\{ 3,4\right\} }^{\prime}\vert\mathbf{X}_{\left\{ 3,4\right\} }\big)=0.5$.
Because the system as a whole is deterministic, $H\left(\mathbf{X}^{\prime}\middle\vert\mathbf{X}\right)=0$
and the total stochastic interaction of partition 12/34 is $H\big(\mathbf{X}_{\left\{ 1,2\right\} }^{\prime}\vert\mathbf{X}_{\left\{ 1,2\right\} }\big)+H\big(\mathbf{X}_{\left\{ 3,4\right\} }^{\prime}\vert\mathbf{X}_{\left\{ 3,4\right\} }\big)-H\left(\mathbf{X}^{\prime}\middle\vert\mathbf{X}\right)=0.5$.

Unfortunately, stochastic interaction is not a suitable cost function
for identifying modular partitions of a multivariate dynamical system
(similarly for multi-information and multivariate non-dynamical systems).
In any such system, a minimal stochastic interaction of 0 will be
assigned to the total partition $\pi_{0}$, and generally a partition
will never have a greater stochastic interaction than any of its refinements
(where one partition is a \emph{refinement }of another if every block
of the former is a subset of some block of the latter). Selecting
partitions using stochastic interaction will thus favor partitions
with large blocks, the total partition being a (possibly non-unique)
global minimum.

Due to this, several authors have proposed normalizing factors that
penalize large partitions \citep{conant1972dsc,balduzzi2008integrated}.
However, the derivation and justification of these normalizing terms
is ad hoc. In this work, we approach the problem of identifying modules
from the point of view of statistical prediction. This yields principled
penalization terms for large partitions and leads us to uncover modular
decompositions with clear interpretations in terms of statistical
modeling.

\section{Statistical modeling and modular decompositions\label{sec:Dynamic-modularity-and-prediction}}

Information theory is intimately connected with statistical modeling
\citep{rissanen2007information}. For example, assume a model that
assigns a probability value to data $\mathbf{x}$:
\begin{equation}
Q\left(\mathcal{\mathbf{x}}\right)=\int_{\Theta}Q\left(\mathcal{\mathbf{x}}\middle\vert\theta\right)\omega\left(\theta\right)\mathrm{d}\theta\label{eq:marginallikelihood}
\end{equation}
This term, called the \emph{marginal likelihood} in the Bayesian literature,
is the expectation of the likehood function $Q\left(\mathcal{\mathbf{x}}\middle\vert\theta\right)$
with respect to distribution $\omega\left(\theta\right)$ over parameter
values.

$Q\left(\mathcal{\mathbf{x}}\right)$ is a measure of predictive fit
to data, and its logarithm is often maximized over\emph{ }parameter
distributions or model choices. Equivalently, one can minimize the
negative of its logarithm, a measure of predictive error called \emph{log
loss}. If data samples are drawn from some true probability distribution
$P\left(\mathbf{X=x}\right)$, then the expectation of the log loss
of the marginal likelihood is:
\[
-\sum_{\mathbf{x}\in\mathbf{X}}P\left(\mathbf{x}\right)\log Q\left(\mathbf{x}\right)=\mathrm{KL}\left(P\middle\Vert Q\right)+H\left(P\left(\mathbf{X}\right)\right)
\]
The KL term (from eq. \ref{eq:kl}) is non-negative, and reaches its
minimum of 0 when the model is perfectly fit, i.e. $Q=P$. It is a
measure of excess prediction error of the model above the minimum
possible. This minimum is specified by the entropy term, and depends
only on the true distribution $P\left(\mathbf{X}\right)$ and not
on model or parameter choices. 

A similar situation holds in the dynamic setting. We call \emph{dynamic
models }those that generate conditional distributions of multivariate
future states $\mathbf{x}^{\prime}$ given starting states $\mathbf{x}$:
\[
Q\left(\mathcal{\mathbf{x}}^{\prime}\middle\vert\mathbf{x}\right)=\int_{\Theta}Q\left(\mathbf{x}^{\prime}\middle\vert\mathbf{x},\theta\right)\omega\left(\theta\right)\mathrm{d}\theta
\]
We look at statistical prediction of dynamical systems from the perspective
of an agent who does not possess a perfectly fit model, but must learn
a dynamic model given previous observations. The agent is provided
with a set of factorized models: for each partition of system variables
$\pi$, there is a dynamic model $Q_{\pi}$ whose parameters and marginal
likelihood obey the independence conditions imposed by the block structure
of $\pi$:
\begin{eqnarray}
Q_{\pi}\left(\mathbf{x}^{\prime}\middle\vert\mathbf{x}\right) & = & \prod_{B\in\pi}Q_{\pi}\left(\mathbf{x}_{B}^{\prime}\middle\vert\mathbf{x}_{B}\right)\label{eq:ind-ml}
\end{eqnarray}

The predictive performance of our agent depends on the chosen model
and the amount of previously observed data. It can be quantified with
a \emph{risk function}, which here is the KL divergence between the
true distribution $P\left(\mathbf{X}^{\prime}\vert\mathbf{X}\right)$
and the distribution predicted by a dynamic model \citep{haussler1997mutual}.
The risk of model $Q_{\pi}$ on the next sample, after observing $N$
previous samples, is: 
\begin{align}
r_{N,Q_{\pi}} & =\mathrm{KL}\left[P\left(\mathbf{X}^{\prime}\middle\vert\mathbf{X}\right)\middle\Vert Q_{\pi}\left(\mathbf{X}^{\prime}\middle\vert\mathbf{X},\mathbf{X}^{\prime1..N},\mathbf{X}^{1..N}\right)\right]\label{eq:instrisk}
\end{align}
The expectation in the KL term is taken over the next sample of $\mathbf{X}^{\prime}\!,\mathbf{X}$,
as well as $N$ previous i.i.d. samples $\mathbf{X}^{\prime1..N}\!,\mathbf{X}^{1..N}$.
The Bayesian \emph{posterior predictive distribution}:
\[
Q_{\pi}\!\left(\mathbf{x}^{\prime}\middle\vert\mathbf{x},\!\mathbf{x}^{\prime1..N}\!,\!\mathbf{x}^{1..N}\!\right)\!=\!\int\negthickspace Q_{\pi}\!\left(\mathbf{x}^{\prime}\middle\vert\mathbf{x},\!\theta\right)Q_{\pi}\!\left(\theta\middle\vert\mathbf{x}^{\prime1..N}\!,\!\mathbf{x}^{1..N}\!\right)\mathrm{d}\theta
\]
is the marginal likelihood of eq. \ref{eq:marginallikelihood}, with
the distribution over parameter values conditioned on $N$ previous
data samples. From the point of view of machine learning, such Bayesian
updating of parameters in light of observed data corresponds to model
\emph{training}, while evaluating the expected model risk on new samples
corresponds to model \emph{testing}. More concretely, our dynamic
models can be considered \emph{supervised learners}: given data, they
infer probabilistic mappings from inputs (starting states $\mathbf{X}$)
to outputs (future states $\mathbf{X^{\prime}}$). 

Given the independence assumption of eq. \ref{eq:ind-ml}, risk $r_{N,Q_{\pi}}$
becomes:
\[
\mathcal{I}_{\pi}\!\left(\mathbf{X}^{\prime}\middle\vert\mathbf{X}\right)+\!\sum_{B\in\pi}\!\mathrm{KL}\!\left[P\!\left(\mathbf{X}_{B}^{\prime}\middle\vert\mathbf{X}_{B}\right)\middle\Vert Q_{\pi}\!\left(\mathbf{X}_{B}^{\prime}\!\middle\vert\mathbf{X}_{B},\!\mathbf{X}_{B}^{\prime1..N}\!,\!\mathbf{X}_{B}^{1..N}\!\right)\!\right]
\]
This form draws attention to the two components that contribute to
risk (that is, predictive error). The \emph{stochastic interaction
term} (see also eq. \ref{eq:stochastic_interaction}) arises as a
consequence of ignoring dynamic coupling between variables in different
blocks. It is the minimal excess error of a factorized\emph{ }model
(in which the dynamics of the variable blocks\textbf{ }induced by
partition $\pi$ are independent) above an optimally fit whole-system
model (where interactions between all variables can be captured).

The second term, called the \emph{complexity term}, reflects the excess
predictive error of a trained model above the minimum possible. It
arises because a model trained on a finite amount of data maintains
some uncertainty about optimal parameter values. For a given amount
of training data, complex models (with larger parameter spaces) will
have greater parameter uncertainty than simpler models, resulting
in more excess predictive error. As $N\rightarrow\infty$, the complexity
term can be asymptotically approximated by $\frac{d_{\pi}}{2N}$,
where $d_{\pi}$ refers to the number of parameters of model $Q_{\pi}$
\citep{komaki1996asymptotic,barron1998information}. This yields:%
\footnote{This approximation assumes continuously-parameterized models and standard
regularity conditions. It also assumes that, for all $\pi$, some
parameterization of $Q_{\pi}$ offers a perfect fit to the factorized
$\Pi_{B\in\pi}P\left(\mathbf{X}_{B}^{\prime}\vert\mathbf{X}_{B}\right).$
It is possible to generalize beyond this case, where the factorizations
of the true distribution are {}`out-of-class' of the models $Q_{\pi}$.%
}
\begin{equation}
r_{N,Q_{\pi}}\approx\mathcal{I}_{\pi}\left(\mathbf{X}^{\prime}\middle\vert\mathbf{X}\right)+\frac{d_{\pi}}{2N}\label{eq:Riskapprox}
\end{equation}

For a given amount of training data $N$, the model with the lowest
risk, 
\[
Q^{\star}\left(N\right)=\arg\min_{Q_{\pi}}r_{N,Q_{\pi}}
\]
corresponds to the partition providing an optimal predictive decomposition
of the system. Models that minimize risk offer a balance between two
conflicting constraints: on one hand, low stochastic interaction (better
predictions under optimal fit), on the other, low model complexity
(easier parameter estimation with limited training data). Because
partitions with smaller blocks (which have smaller-state-space dynamics
representable by fewer parameters) generally induce simpler models,
risk presents a principled cost function for identifying small, weakly-coupled
modules. The amount of data $N$ parameterizes this trade-off: as
$N$ increases, emphasis is shifted from the complexity term to the
stochastic interaction term, and groups of variables whose dynamic
interactions carry the most information while being easiest to learn
are first to coalesce into multivariate blocks of the optimal model.%
\footnote{Minimizing risk can be seen as a form of information bottleneck \citep{tishby1999ibm}:
it searches for factorized models whose parameters minimize information
about training data while maximizing information about system dynamics;
the size of the training data serves as a trade-off parameter.%
} Thus, selecting optimal decompositions while increasing the amount
of training data generates a modular multiscale decomposition of system
variables. In the infinite data limit, the risk of each model $Q_{\pi}$
reaches its minimum of $\mathcal{I}_{\pi}\left(\mathbf{X}'\vert\mathbf{X}\right)$,
and the partition corresponding to $Q^{\star}$ becomes the one with
lowest stochastic integration (the total partition being a possibly
non-unique minimum).

\section{Decomposing a dynamical system}

The complexity term in eq. \ref{eq:Riskapprox} depends on the parametric
form of the dynamic model. Though a variety of possibilities exist,
here our dynamic models are assumed to be products of first-order
Markov chains with Dirichlet priors. The number of parameters of model
$Q_{\pi}$ from this class is:
\begin{equation}
d_{\pi}=\sum_{B\in\pi}\left|\mathbf{X}_{B}\right|\left(\left|\mathbf{X}_{B}^{\prime}\right|-1\right)\label{eq:numparams}
\end{equation}
where $\left|\mathbf{X}_{B}\right|$ is the number of supported starting
state outcomes and $\left|\mathbf{X}_{B}^{\prime}\right|$ is the
number of possible future state outcomes of the variables with indexes
in block $B$. For example, for a single block of Boolean variables
with a fully supported starting state distribution, these are both
equal to $2^{\left|B\right|}$. For this model class, the complexity
term scales exponentially with the number of variables in each block.

\begin{figure}[t]
\includegraphics[bb=0bp 15bp 429bp 570bp,clip,width=1\columnwidth]{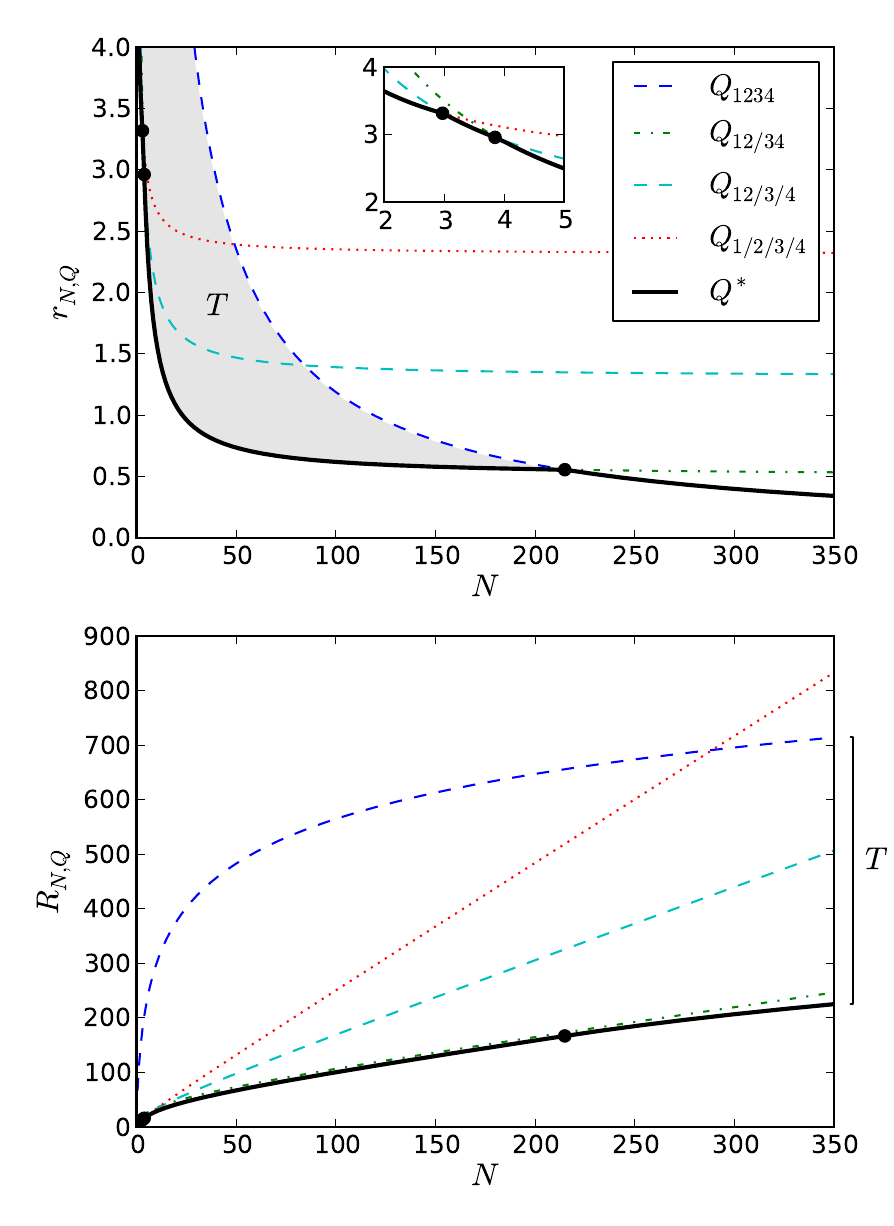}\caption{Top: approximate risk for optimally-predictive models of the Boolean
network from fig.  \ref{fig:Simple-four-node}. Dots mark switches
of the optimal model $Q^{\star}$; inset shows first two switches.
Bottom: cumulative risk, or total accumulated prediction error for
models plotted in the top graph. Total modularity ($T$) is asymptotic
difference between cumulative risks of $Q_{1234}$ and $Q^{\star}$
or, alternatively, area between lines corresponding to (non-cumulative)
risks of $Q_{1234}$ and $Q^{\star}$. \label{fig:fournoderisk}}
\end{figure}

As an example, we look at optimal decompositions of the network in
fig. \ref{fig:Simple-four-node}. Its risk, calculated using the approximation
of eq. \ref{eq:Riskapprox} and parameter counts of eq. \ref{eq:numparams},
is shown at the top of fig. \ref{fig:fournoderisk}.%
\footnote{In general, the approximation of eq. \ref{eq:Riskapprox} is only
accurate for large $N$. However, it suffices for our explanatory
purposes.%
} The risk is plotted for those models which reach minimum risk at
some point of the training process, as well as that of the overall
minimal risk model $Q^{\star}$ at each $N$. Predictive power is
initially optimized by the model corresponding to partition 1/2/3/4
(the simplest model which treats all nodes independently). At $N\approx3$
(inset), it is replaced by the model corresponding to partition 12/3/4
(variables 1 and 2 now merged into a single block); at $N\approx4$
(inset), by the model corresponding to partition 12/34; and finally
at $N\approx215$, the most predictive model becomes the one corresponding
to the total partition 1234.

\section{Total modularity\label{sec:Total-modularity}}

So far, our measure of modularity has been parameterized by $N$,
the amount of training data. Here, we derive a parameter-free measure
of the \emph{total modularity} in a dynamical system. 

In our definition of risk (eq. \ref{eq:instrisk}), we used the \emph{posterior
predictive distribution }$Q_{\pi}\left(\mathbf{X}^{\prime}\middle\vert\mathbf{X},\mathbf{X}^{\prime1..N},\mathbf{X}^{1..N}\right)$,
the probability assigned to the next data sample by a model trained
on $N$ previous data samples. Given our assumptions, the following
relationship holds between the\emph{ prior predictive distribution},
the probability an untrained model assigns to $N$ data samples, and
the posterior predictive distribution:
\[
Q_{\pi}\!\left(\mathbf{X}^{\prime1..N}\middle\vert\mathbf{X}^{1..N}\right)\!=\!\prod_{n=0}^{N-1}\! Q_{\pi}\left(\mathbf{X}^{\prime n+1}\middle\vert\mathbf{X}^{n+1}\!,\mathbf{X}^{\prime1..n}\!,\mathbf{X}^{1..n}\right)
\]

This suggests the \emph{prequential }interpretation of Bayesian prediction
\citep{dawid1992prequential}: the expected predictive error of a
model on $N$ samples is the sum of the expected predictive errors
on each successive sample after training on the previous samples.
This accumulated prediction error is termed \emph{cumulative risk}
\citep{haussler1997mutual}:
\[
R_{N,Q_{\pi}}=\sum_{n=0}^{N-1}r_{n,Q_{\pi}}
\]
The risk of eq. \ref{eq:instrisk} can be seen as the rate of change
of the cumulative risk as the amount of training data grows. 

\emph{Total modularity} is the total gain in predictive accuracy (i.e.,
decrease in cumulative risk) provided by the optimally predictive
models $Q^{\star}\left(N\right)$ versus the unfactorized, total-partition
model $Q_{\pi_{0}}$. Let $R_{N,Q^{\star}}=\sum_{n=0}^{N-1}r_{n,Q_{\pi}^{\star}\left(n\right)}$
be the cumulative risk of an agent who selects the risk-minimal model
at each $N$. The total modularity is then:
\begin{equation}
T=\lim_{N\rightarrow\infty}\left(R_{N,Q_{\pi_{0}}}-R_{N,Q^{\star}}\right)\label{eq:totalmod}
\end{equation}

Total modularity measures the overall predictive advantage gained
by using factorized models, and is not a function of a particular
$N$. High values of total modularity indicate that simpler models
have significantly improved predictive performance during earlier
stages of the learning process\emph{.}%
\footnote{Minimization of accumulated error by online switching from simpler
to more complex models is related to a learning framework recently
proposed by \citet{van2007catching}%
} To use the previous example, the cumulative risk of the models plotted
at the top of fig. \ref{fig:fournoderisk} is shown at the bottom
of that figure. The total modularity of the dynamic network shown
in fig. \ref{fig:Simple-four-node} is equal to the asymptotic difference
between the cumulative risks of $Q_{1234}$ ($=Q_{\pi_{0}}$) and
$Q^{\star}$. Equivalently, it is also the total area between the
lines corresponding to the (non-cumulative) risks of $Q_{1234}$ and
$Q^{\star}$.

\begin{figure}
\includegraphics[bb=0bp 5bp 305bp 185bp,clip,width=1\columnwidth]{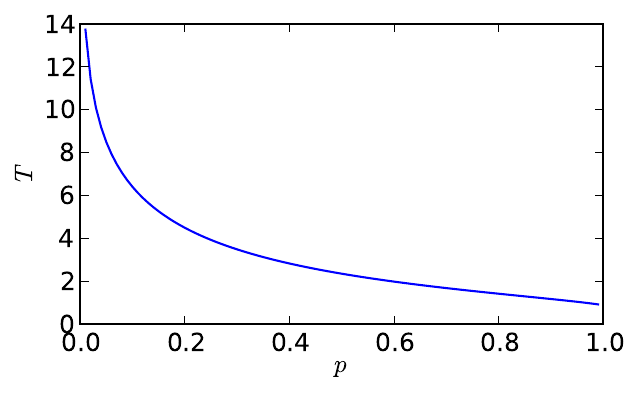}

\caption{Total modularity of two binary variables which copy each others'
state with probability $p$ and maintain their own state with probability
$1-p$. Total modularity increases as coupling decreases, and diverges
as $p\rightarrow0$.\label{fig:switchrisk}}
\end{figure}

For another illustration of total modularity, we consider a simple
dynamical system composed of two binary variables. Each variable is
parameterized in the following manner: at each time step, with probability
$p$ it assumes the value of the other variable in the previous time
step, and with probability $1-p$ it maintains its own value from
the previous time step. The amount of dynamic coupling between the
two nodes increases with $p$: at $p=0$ the variables have no interaction,
while at $p=1$ their values are completely correlated (with a one
timestep lag). This dynamic coupling is illustrated in fig. \ref{fig:switchrisk},
which plots the total modularity of this system against the coupling
parameter $p$. The total modularity monotonically decreases as $p$
increases, showing that greater coupling leads to lower total modularity.
As $p\rightarrow0$, the two variables become completely independent
and total modularity diverges (in this case, it grows without bound
at a rate proportional to $\log N$).

\section{State-dependent and causal modularity\label{sec:Causal-modularity}}

The way information flows within a dynamical system can depend on
the system's state. For example, a partition's stochastic interaction
can be different in different attractors. We can quantify this by
different choices of the starting state distribution, $P\left(\mathbf{X}\right)$.
Though we have generally taken $P\left(\mathbf{X}\right)$ to be a
fully-supported uniform distribution, it can be weighted preferentially
over some subset of starting states.

\begin{figure}[t]
\includegraphics[width=1\columnwidth]{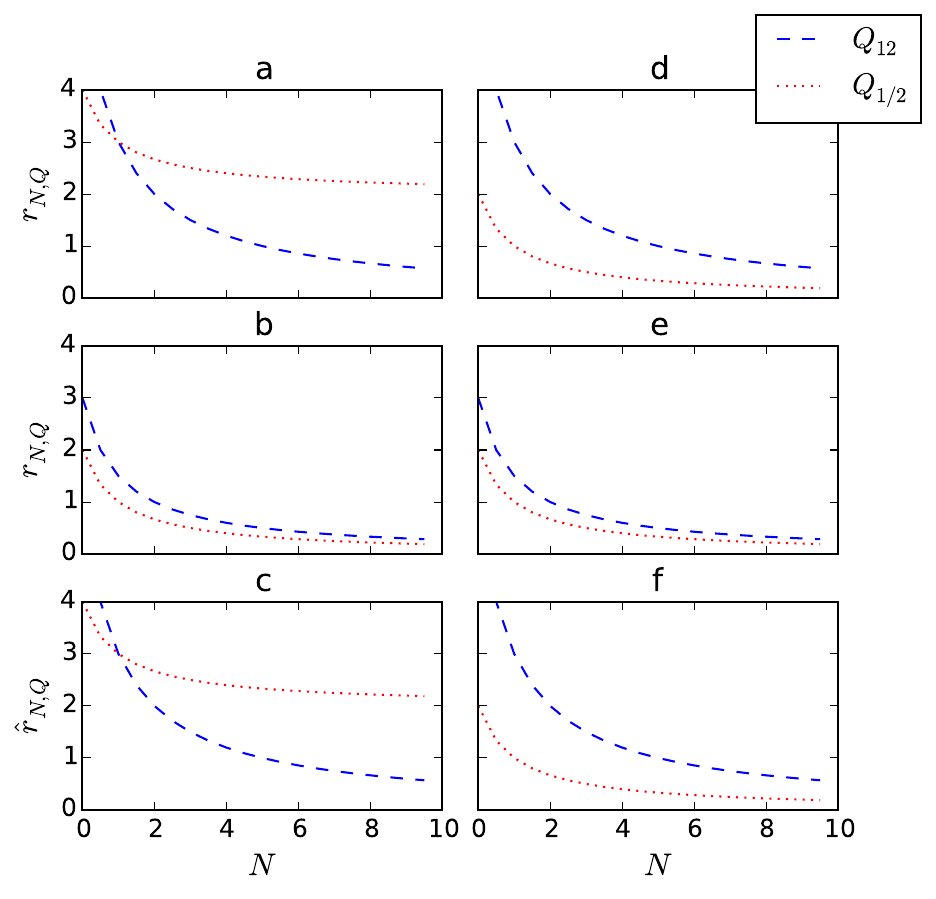}

\caption{Risk for two systems, each having two binary variables: in system
$A$ (left column) each variable copies previous value of the other,
in system $B$ (right column) each variable takes opposite of its
own previous state. a) and d):\textbf{ }Risk under uniform starting
state distribution. Lowest risk model of $A$ becomes the total one,
while factorized model remains optimal for $B$. b) and e): Risk and
optimal decompositions depend on the starting state distribution.
Computed over $P\left(\mathbf{X}\!=\!\left(0,1\right)\right)\!\!=\!0.5,P\left(\mathbf{X}\!=\!\left(1,0\right)\right)\!\!=\!0.5$,
risk and optimal decompositions become the same for $A$ and $B$,
though their causal organization is different. c) and f): \emph{Causal
risk} leads to different decompositions of $A$ and $B$, even when
computed over same starting state distribution as in b) and e).\label{fig:causalrisk}}
\end{figure}

For example, consider two systems, each composed of two binary variables.
In system $A$, each variable copies the previous value of the other,
while in system $B$, each variable takes the opposite of its own
previous state. Fig. \ref{fig:causalrisk} shows the risk plots for
both $A$ (left column) and $B$ (right column), where \ref{fig:causalrisk}a
and \ref{fig:causalrisk}d are calculated for a uniform starting state
distribution. The risk, as well as the optimal decompositions, is
different between the two systems: $A$ (which performs the copy operation)
eventually chooses the total partition $\left\{ \left\{ 1,2\right\} \right\} $
as the most predictive, while $B$ (whose variables perform independent
state flips) never does.

If, however, a non-uniform starting state distribution is chosen,
risk and optimal decompositions can change. The risk for starting
state distribution $P\left(\mathbf{X}\!=\!\left(0,1\right)\right)\!=\!0.5,P\left(\mathbf{X}\!=\!\left(1,0\right)\right)\!=\!0.5$
are shown in fig. \ref{fig:causalrisk}b and \ref{fig:causalrisk}e
(for systems $A$ and $B$ respectively). Different parts of the starting
state space induce different risk values and optimal decompositions:
for this distribution, fig. \ref{fig:causalrisk}b shows that the
total partition $\left\{ \left\{ 1,2\right\} \right\} $ is never
chosen as the optimally predictive one for system $A$.

Additionally, for these starting states the transition distributions
of $A$ and $B$ are identical: if either system is started in state
$\left(0,1\right)$, it deterministically transitions to state $\left(1,0\right)$,
and similarly for the transition from $\left(1,0\right)$ to $\left(0,1\right)$.
Because the observed dynamics of the two systems are identical, the
risk functions and optimal decompositions are also equal. Though systems
$A$ and $B$ are defined using different causal architectures, here
their modular organizations are indistinguishable. Specifically, $A$
is postulated to have a causal connection among its variables but
-- for this starting state distribution -- they display no stochastic
interaction.

This example highlights the difference between statistical correlation
and causal interaction. To properly handle the latter, we utilize
a notion of causality based on semantics of intervention \citep{pearl2000causality},
recently developed in an information-theoretic direction by \citet{ay2008information}.
In Pearl's treatment, conditional probability distributions represent
not only correlations, but also responses of variables to externally-imposed
interventions. This is especially natural when dynamics of interest
are generated by causal models, such as dynamic causal Bayesian or
Boolean network models frequently used in artificial life and systems
biology.

In our example, the functional organization of systems $A$ and $B$
can be differentiated -- even within the non-uniform starting state
distribution mentioned above -- if the starting states of the systems
can be intervened upon. This is because in system $A$ -- but not
system $B$ -- changing the starting state of one variable can change
the other variable's future state.

We consider interventions formally by noting that the risk $r_{N,Q_{\pi}}$
of eq. \ref{eq:instrisk} need not take the same starting state distribution
for training data as for the testing data. Instead, we take the starting
state distribution for training data to be drawn i.i.d. from a fully-supported
and uniform distribution $\hat{P}\big(\mathbf{X}\big)$ (the distribution
of interventions), while the testing starting states can be drawn
from any $P\left(\mathbf{X}\right)$ of interest. We refer to risk
evaluated under this learning scenario as \emph{causal risk}:
\begin{multline*}
\!\!\!\!\!\!\hat{r}_{N,Q_{\pi}}=\sum_{\mathbf{x},\mathbf{x}^{\prime}}\! P\left(\mathbf{x}\right)P\left(\mathbf{x}^{\prime}\middle\vert\mathbf{x}\right)\Big[\log P\left(\mathbf{x}^{\prime}\middle\vert\mathbf{x}\right)-\\
\sum_{\mathbf{x}^{1\!.\!.\! N}\!\!,\mathbf{x}^{\prime1\!.\!.\! N}}\!\!\!\!\!\!\hat{P}\left(\mathbf{x}^{1..N}\!\right)P\left(\mathbf{x}^{\prime1..N}\!\middle\vert\mathbf{x}^{1..N}\!\right)\!\log Q_{\pi}\!\left(\mathbf{x}^{\prime}\middle\vert\mathbf{x},\!\mathbf{x}^{\prime1..N}\!,\!\mathbf{x}^{1..N}\!\right)\Big]\!\!\!\!\!\!\!
\end{multline*}
As $N\rightarrow\infty$, the posterior predictive distribution of
model $Q_{\pi}$ approaches $\prod_{B\in\pi}\hat{P}\big(\mathbf{X}_{B}^{\prime}\vert\mathbf{X}_{B}\big)$,
where $\hat{P}\big(\mathbf{X}_{B}^{\prime}\vert\mathbf{X}_{B}\big)$
is the whole-system transition distribution $P\big(\mathbf{X}^{\prime}\vert\mathbf{X}\big)$
marginalized onto variables in block $B$ using $\hat{P}\big(\mathbf{X}\big)$.
Then, $\hat{r}_{N,Q_{\pi}}$ can be approximated by:
\[
\mathcal{I}_{\pi}\left(\mathbf{X}^{\prime}\middle\vert\mathbf{X}\right)+\sum_{B\in\pi}\mathrm{KL}\left[P\left(\mathbf{X}_{B}^{\prime}\middle\vert\mathbf{X}_{B}\right)\middle\Vert\hat{P}\left(\mathbf{X}_{B}^{\prime}\middle\vert\mathbf{X}_{B}\right)\right]+\frac{\hat{d}_{\pi}}{2N}
\]
where $\mathcal{I}_{\pi}$ and the expectations in the
KL terms use the testing starting state distribution, while $\hat{d}_{\pi}$ is the number of training parameters. The KL divergence
between $P\big(\mathbf{X}_{B}^{\prime}\vert\mathbf{X}_{B}\big)$ (the
whole-system transition distribution marginalized onto variables in
block $B$ using $P\big(\mathbf{X}\big)$ ) and $\hat{P}\big(\mathbf{X}_{B}^{\prime}\vert\mathbf{X}_{B}\big)$
reflects the amount of extra perturbation that active interventions
inject into block dynamics. The two distributions need not be equal,
unless $P\left(\mathbf{X}\right)=\hat{P}\big(\mathbf{X}\big)$ or
the partition under consideration is the total one. Because KL divergence
is non-negative, causal risk $\hat{r}_{N,Q_{\pi}}$ is not less than
the statistical risk $r_{N,Q_{\pi}}$(compare above to eq. \ref{eq:Riskapprox}).

Fig. \ref{fig:causalrisk}c and \ref{fig:causalrisk}f show the causal
risk for systems $A$ and $B$ (respectively) with $P\left(\mathbf{X}\!=\!\left(0,1\right)\right)\!=\!0.5,P\left(\mathbf{X}\!=\!\left(1,0\right)\right)\!=\!0.5$.
In \ref{fig:causalrisk}c -- but not \ref{fig:causalrisk}f -- the
total partition model assumes a lower risk than the factorized model,
indicating that for the starting states in question, system $A$ --
but not system $B$ -- has causal interactions between its variables.

\section{Conclusion}

Modularity is normally treated as an objective property of a system's
organization. Our approach instead considers from the perspective
of modeling and prediction. In the context of inferring dynamic models
from limited data, modularity allows for models that are predictive
but simple, with the amount of training data controlling the trade-off.
Our statistical treatment connects to previous information-theoretic
approaches, but goes further by providing principled terms for identifying
small modules.

Our approach can also be used to find state-dependent modular organizations,
both in statistical and causal (interventional) senses: models trained
on interventional dynamics but tested on arbitrary distributions give
rise to a measure that identifies causal modules. This is related
to existing information-theoretic measures of causal interactions
between subsystems \citep{tononi2003measuring}, but here emerges
naturally from the framework of statistical modeling. This framework
also produces a measure of total modularity present in the system,
which quantifies the overall predictive advantage that modularity
provides through the entire model inference process.

As a side note, if the learning of real-world cognitive systems (such
as scientists or organisms) proceeds in a manner somewhat similar
to the statistical framework presented here, our approach suggests
why such systems may infer modular organizations in the external world:
under conditions of limited data, this assumption can simplify learning
and lead to gains in predictive power.

One important issue with our treatment is its model-dependence. The
complexity penalization term of eq. \ref{eq:instrisk} depends on
the model class, and different model classes may have different parameterizations
and functional forms. Our examples employed products of Markov chain
models, a rather general dynamic model class but one heavily parameterized;
others could be used. The choice of model class can be thought of
as a null model of system dynamics.  

Several generalizations suggest themselves. For example, it is possible
to infer module timescales by searching not only over decompositions,
but also model orders (numbers of previous states on which transition
probabilities depend; for inferring Markov chain order, see \citealp{strelioff2007inferring}).
Fuzzy modular organizations, in which a variable can belong to more
than one module, can be accommodated by allowing partially-overlapping
blocks. More generally, the model search space could include other
structures besides partitions (e.g. trees or networks) to impose independence
constraints on information flow between blocks.

Identifying modularity in dynamical systems is important in complex
systems research in general, and biological systems modeling in particular.
Our method differs from recent community-detection methods that find
modularity in static graphs, in that it focuses on the organization
of interactions between dynamic system components. In future work,
we hope to apply it to the analysis of regulatory and signaling control
in biochemical networks, as well as inference of functional neural
organization from brain recordings.

\section{Acknowledgments}

Thanks to Randy Beer, Paul Williams, Olaf Sporns, and the participants
of the \emph{Guided Self-Organization 3} workshop for useful feedback
and encouragement.

\footnotesize

\bibliographystyle{apalike}
\bibliography{KOLCHINSKY}

\end{document}